\documentclass[twocolumn,aps,showpacs,prl]{revtex4-1}
\usepackage{amsmath,amssymb}
\usepackage{subfig}
\usepackage[countmax]{subfloat}
\usepackage{graphicx}
\usepackage[usenames]{color}
\usepackage{hyperref}

\begin{document}

\title{Universal power law in crossover from integrability to quantum chaos}

\author{ Ranjan Modak$^1$, Subroto Mukerjee$^{1,2}$ and
Sriram Ramaswamy$^{1,3}$}
\affiliation{$^1$ Department of Physics, Indian Institute of Science, Bangalore 560 012, India}
\affiliation{$^2$ Centre for Quantum Information and Quantum Computing, Indian Institute of Science,
Bangalore 560 012, India}
\affiliation{$^3$ TIFR Centre for Interdisciplinary Sciences, Hyderabad 500 075,
India}

\begin{abstract}
We study models of interacting fermions in one dimension to investigate the
crossover from integrability to non-integrability, i.e., quantum chaos, as
a function of system size. Using exact diagonalization of finite-sized systems,
we study this crossover by obtaining the energy level statistics and Drude
weight associated with transport. Our results reinforce the idea that for
system size $L \to \infty$ non-integrability sets in for an arbitrarily small
integrability-breaking perturbation. The crossover value of the perturbation
scales as a power law $\sim L^{-3}$ when the integrable system is gapless and
the scaling appears to be robust to microscopic details and the precise form of
the perturbation. We conjecture that the exponent in the power law is
characteristic of the random matrix ensemble describing the non-integrable
system. For systems with a gap, the crossover scaling appears to be faster than
a power law.
\end{abstract}


\maketitle

How isolated quantum systems thermalize, hitherto investigated
theoretically in a few special
cases~\cite{deutsch1991,srednicki1994,rigol.2008}, is now the subject of
active experimental study thanks to the advent of cold-atom
systems~\cite{kinoshita2006,Gring14092012}.
Recall that in isolated classical
systems that thermalize, a phase space trajectory samples all possible
microstates at a given energy spending equal amounts of time in each,
yielding the microcanonical prescription. On the
other hand, for a system which does not thermalize, the trajectory typically
follows regular, not chaotic, orbits constrained by conservation laws and
samples only a low-dimensional subspace.
This notion of thermalization underpins the Fermi-Pasta-Ulam problem of a
classical system of masses connected by springs~\cite{fermi1955}. For
harmonic springs the system does not thermalize and, even upon the introduction
of anharmonicity, thermalization occurs only above an energy threshold
which, however, scales to zero with increasing system size, as a power-law
characterizing the nature of anharmonicity~\cite{pettini.1990}. Signatures of
lack of thermalization in classical systems can also be seen in
transport~\cite{dhar2008heat} (but note that singular size-dependence of
thermal conductivity is distinct from a failure to
thermalize~\cite{ramaswamy2002}).

In this paper we investigate analogous issues for quantum systems.
Later in the paper we compare our study to the related work of Rabson {\em
et al.}~\cite{rabson.2004}. As quantum mechanics lacks a notion of phase space,
we identify thermalization with non-integrability, i.e., quantum
chaos, a now-standard prescription,
and use the corresponding diagnostic tools.
We would like to emphasize that
while the integrable systems we study a) are exactly solvable, b)
have an infinity of conservation laws in the thermodynamic limit and c)
display Poissonian level-spacing statistics it is
perhaps only the
last one that is important to prevent thermalization:
localized phases of disordered systems lacking
properties a) and b) have been argued to not thermalize~\cite{huse.2010}.

Our main result is that the characteristic value of control parameter at which significant
non-integrability is seen scales to zero with increasing system size as in the classical
systems described above. For gapless systems,
the approach to zero is a power law whose exponent appears robust to microscopic
details such as variations in the type of integrable limit and
integrability-breaking perturbation.
%
We conjecture that this power law is determined only by
the random matrix ensemble describing the non-integrable system. For systems
with a gap our numerics suggest a faster-than-power-law dependence
of the crossover on system size.

We consider two one-dimensional models of interacting fermions with periodic
boundary conditions that have integrable limits. First is the
$t-t'-V-V'$ model of spinless fermions with Hamiltonian
\begin{eqnarray}\nonumber
H & = & -t\sum_i \left(c^\dagger_i c_{i+1} + \rm{h.c.} \right) - t'\sum_i \left(c^\dagger_i c_{i+2} + \rm{h.c.} \right) \\
& & + V\sum_i n_i n_{i+1} + V'\sum_i n_i n_{i+2}.
\end{eqnarray}
\label{Eqn:t-t'-V}
This model, which can be mapped on the spin 1/2 $XXZ$ chain, is integrable and
exactly solvable by the Bethe ansatz when $t'=V'=0$~\cite{takahashi.1972}. The
other Hamiltonian we study is the Hubbard model with next nearest neighbor
hopping and spin dependent hopping given by
\begin{eqnarray}\nonumber
H & = & -\sum_{i\sigma} t_\sigma \left(c^\dagger_{i\sigma} c_{i+1 \sigma} +
\rm{h.c.} \right) - \sum_{i\sigma} t'_\sigma \left(c^\dagger_{i\sigma} c_{i+2
\sigma} + \rm{h.c.} \right) \\
& & + U\sum_i n_{i\uparrow} n_{i\downarrow}.
\label{Eqn:hubbard}
\end{eqnarray}
This reduces to the regular one dimensional Hubbard model for $t'_\sigma=0$ and
$t_\uparrow = t_\downarrow$, which too is integrable and solvable by the Bethe
ansatz~\cite{lieb.1968}. For both of the above models, we set $t=1$ henceforth.

For the $t-t'-V-V'$ model we have investigated the breaking of integrability by
taking one or both of $t'$ and $V'$ to be non-zero~\cite{lea2.2010}.
The only conserved quantities in the non-integrable cases
are particle number and crystal momentum~\footnote{When $t'=0$, the model is
particle-hole symmetric at half filling, so we work away from half filling.}.

Our Hubbard model is non-integrable in the presence of next-nearest neighbor
hopping. In this case, in addition to particle number and crystal momentum, all
components of the total spin are conserved if $t_\uparrow = t_\downarrow$ and
$t'_\uparrow = t'_\downarrow$. To break this symmetry, we choose $t'_\downarrow
=0$ which breaks the $SU(2)$ symmetry of the model and only $S_z$ is conserved.
Further, the $S_z=0$ sector no longer has degeneracies arising from spin
inversion.

We investigate the breaking of integrability through 1) The energy-level
spacing distribution and 2) the Drude weight for charge transport.

{\em Energy level spacing:} An integrable system without disorder has an
infinite number of conserved quantities in the thermodynamic limit whose values
can be used to label the energy eigenstates of the system. The
energy level spacing obtained from symmetry sectors labeled by any finite set of
quantum numbers shows no level
repulsion and in fact obeys a Poissonian distribution $P(s)=\exp(-s),$ for the
energy spacing $s$ in units of the mean level spacing~\cite{mehta}. On the
other hand, a non-integrable system of the type we study has a finite number of
conserved quantities even in the thermodynamic limit. Once these have been
accounted for the resultant symmetry sectors have no degeneracies left and
the energy levels display level repulsion. $P(s)$
then corresponds to that of a random matrix ensemble even though there is no
inherent microscopic randomness in the Hamiltonian. For most of our studies,
the non-integrable system is described by $P(s)=\pi s/2 \exp(-\pi s^2/4)$,
corresponding to the Gaussian Orthogonal Ensemble (GOE). We thus track $P(s)$
as a function of the strength of the appropriate integrability breaking
parameter (say $p$) to locate the crossover from integrable to non-integrable
behavior. Since our goal is to locate this crossover as a function of system
size, we perform numerical exact diagonalization on finite-sized systems to
obtain {\em all} the energy eigenvalues. We are thus restricted to a maximum
system size of about $L=22$ for the model of spinless fermions and about $L=11$
for the Hubbard model~\cite{lea1.2010}. We perform the diagonalization in
momentum space and leave out the $k=0$ and $k=\pi$ sectors to exclude the
effect
of parity symmetry. The system sizes we consider appear sufficient to quantify
the crossover we are investigating. Having obtained $P(s)$ for a given
system, we locate its peak $S$ by fitting to a Brody
distribution~\cite{brody.1981}
\begin{eqnarray}
P(s)=(\beta +1)bs^{\beta}\exp(-bs^{\beta+1}),
\label{Eq:Brody}
\end{eqnarray}
where $b=\Gamma [(\beta+2)/(\beta+1)]^{\beta+1}$,
which interpolates smoothly between Poissonian ($\beta=0$) and
GOE ($\beta=1$). We assume that all our systems become non-integrable in the
limit $p \rightarrow \infty$~\footnote{For our particular microscopic models,
this is not strictly true since they again become integrable in the limit $t'
\rightarrow \infty$. We are assuming that there is some value $t'_{max}(L)$,
where the model is most non-integrable (i.e. is closest to GOE) and that
$t'_{cr}(L)/t'_{max}(L) \ll 1$.}. Thus, knowing $S$ as a function of $p$, for a
system size $L$, we locate the crossover value $p_L$ by fitting the values of
$S$ to a function of $p$ that smoothly interpolates from 0 (Poisson) at $p=0$
to 0.8 as $p \rightarrow \infty$ (GOE). We choose the function $S(p) = 0.8
\tanh \left(p/p_L \right)$~\cite{rabson.2004}for this purpose and have checked
that other functions yield similar results.

\begin{figure}
\centering
\begin{tabular}{cc}
\includegraphics[width=1.7in]{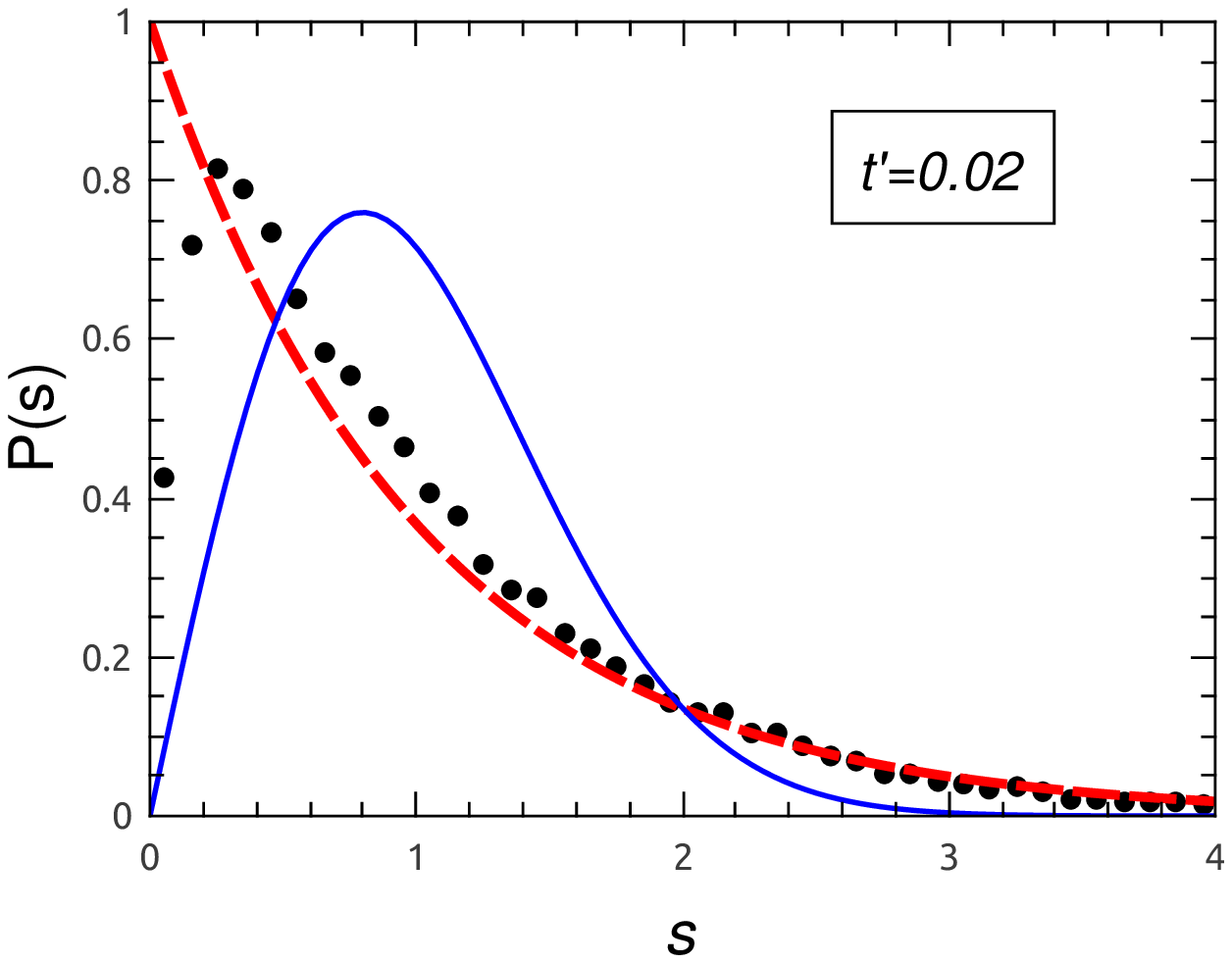} &
\includegraphics[width=1.7in]{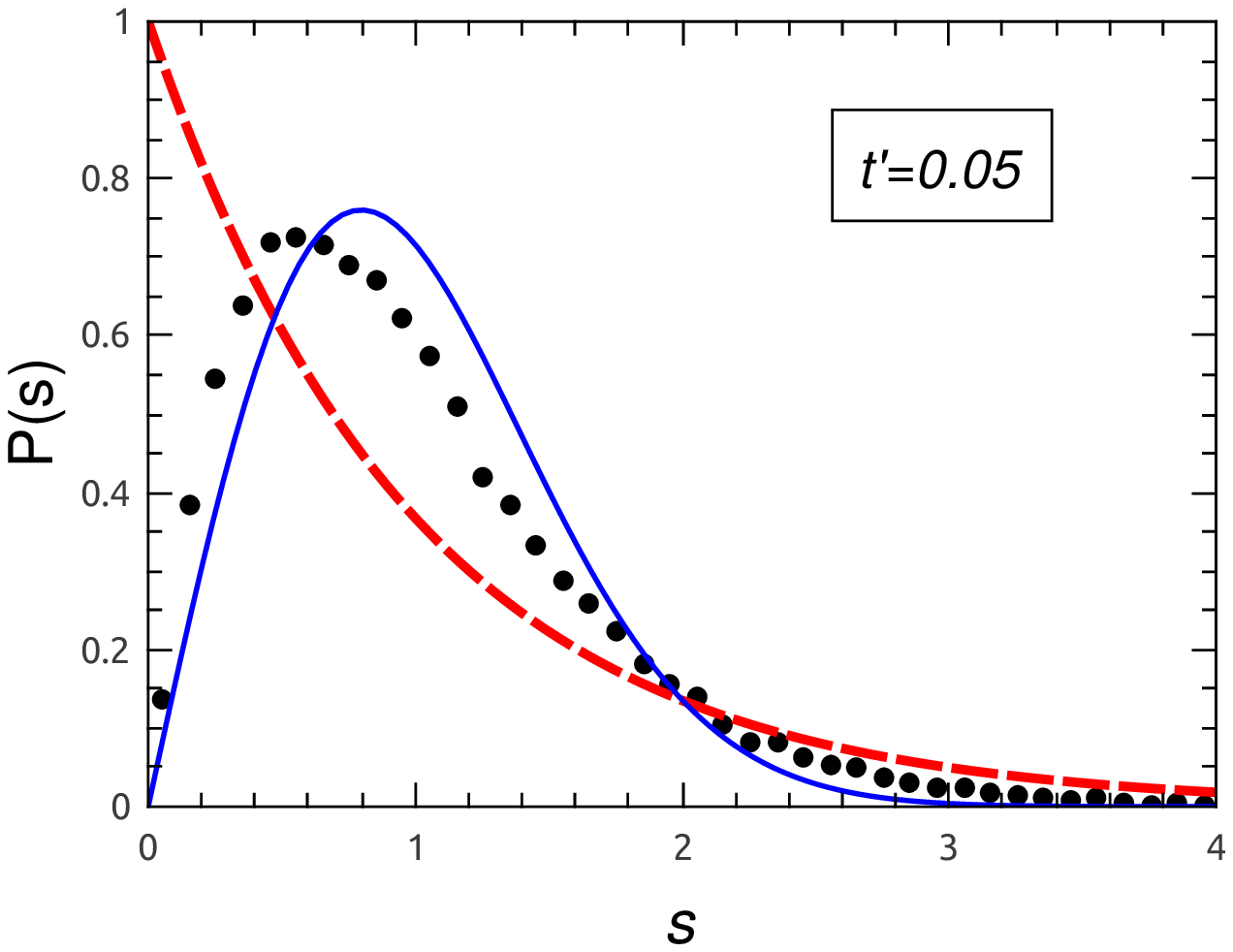} \\
\includegraphics[width=1.7in]{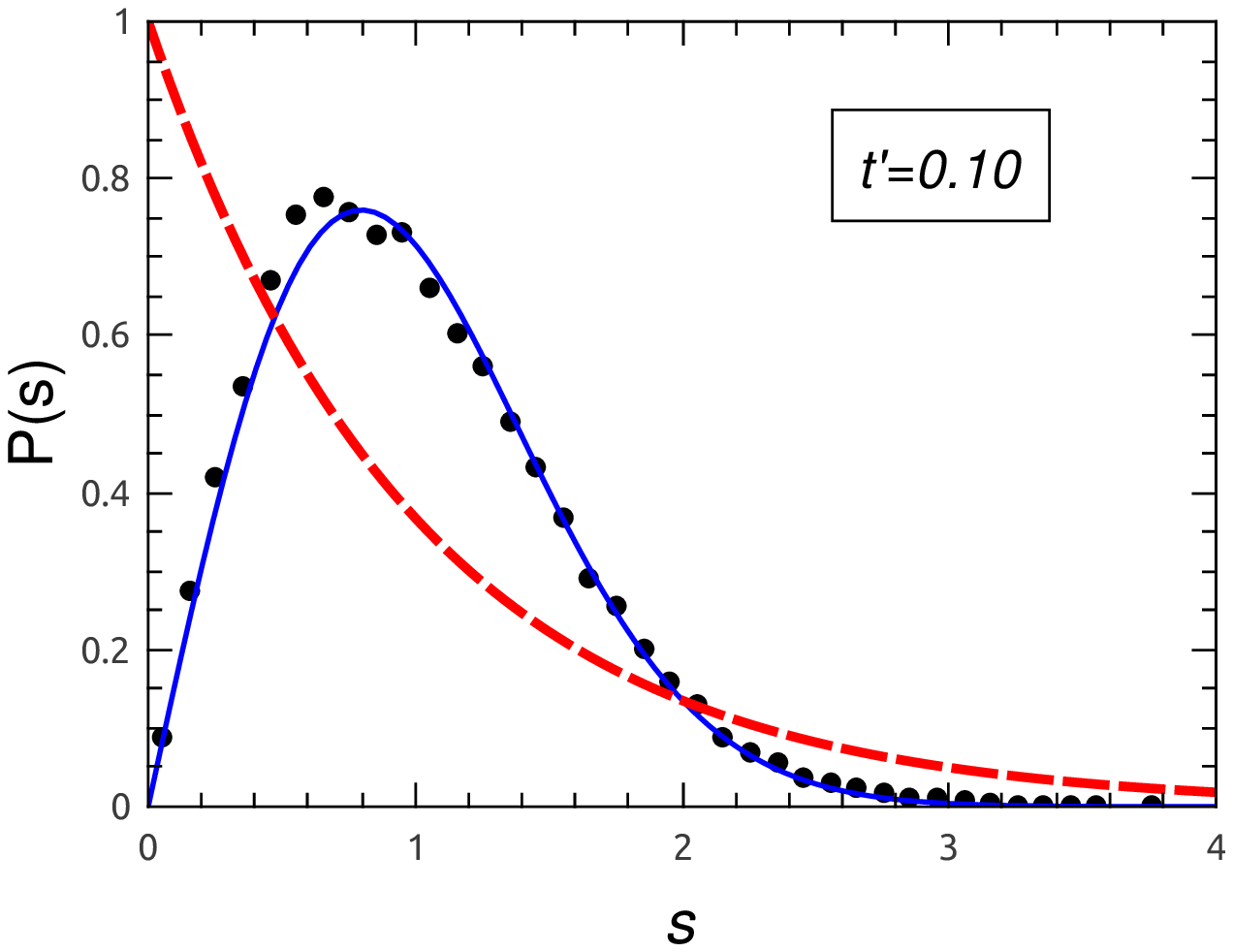} &
\includegraphics[width=1.7in]{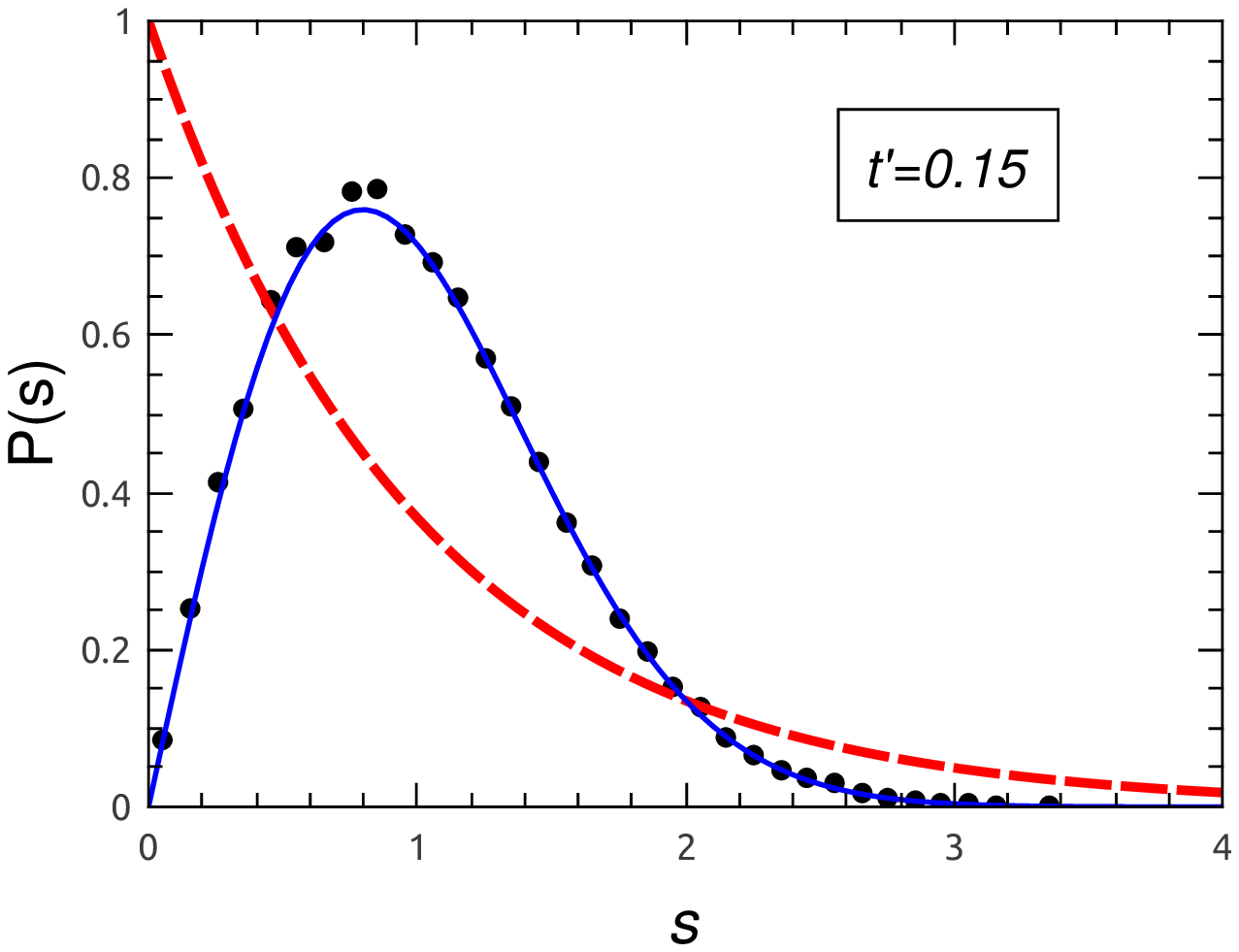}
\end{tabular}
\caption{(Color online) Level spacing distribution $P(s)$ for the $t-t'-V$
model with $V=1$ at half filling and $L=22$. The values of the integrability
breaking parameter $t'$ are 0.02,0.05,0.1,0.15. The dashed line is the Poisson
distribution and the solid line the level spacing distribution for GOE.}
\label{Fig:levelspacing_tv}
\end{figure}

\begin{figure}
\centering
\begin{tabular}{cc}
\includegraphics[width=1.7in]{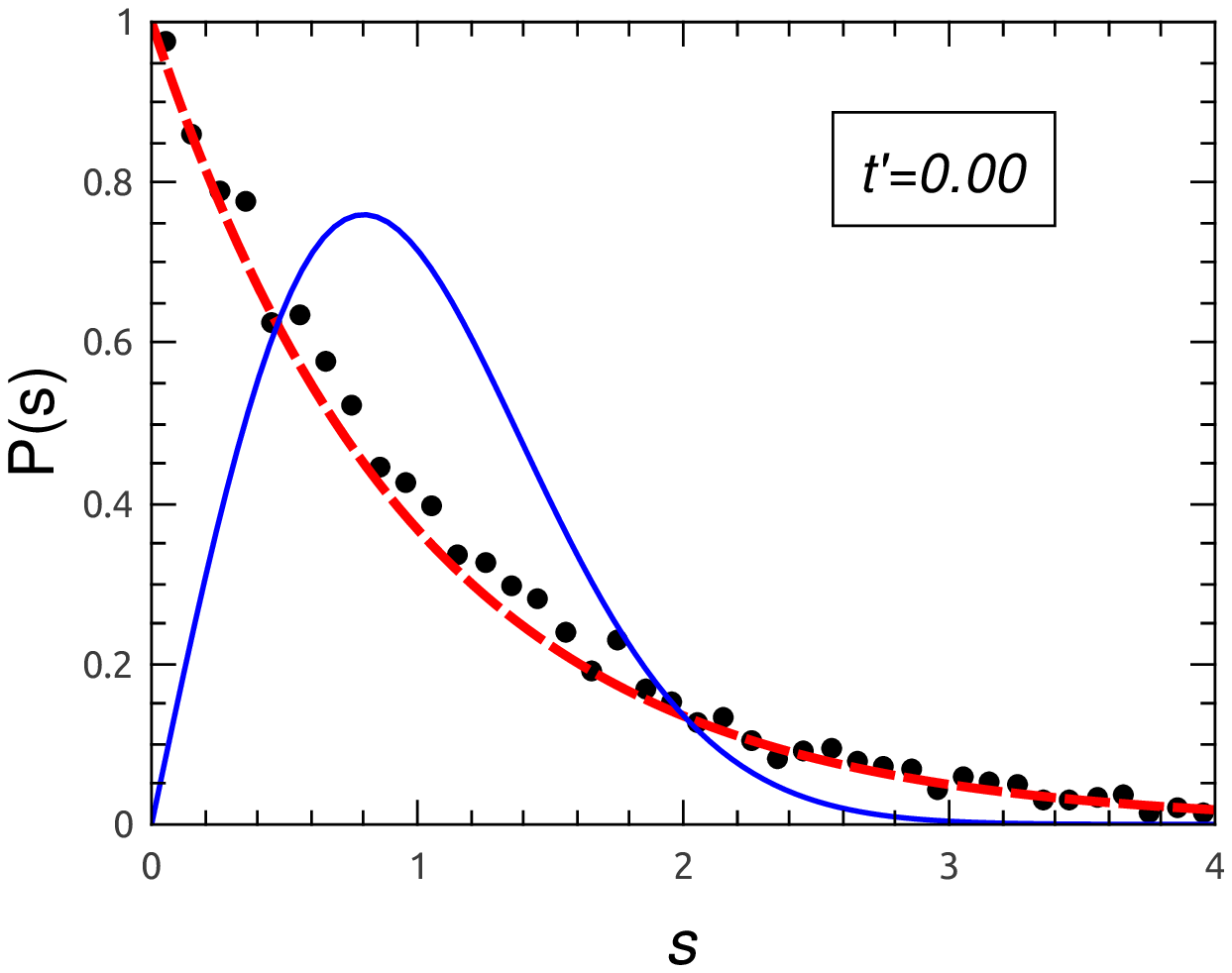}&
\includegraphics[width=1.7in]{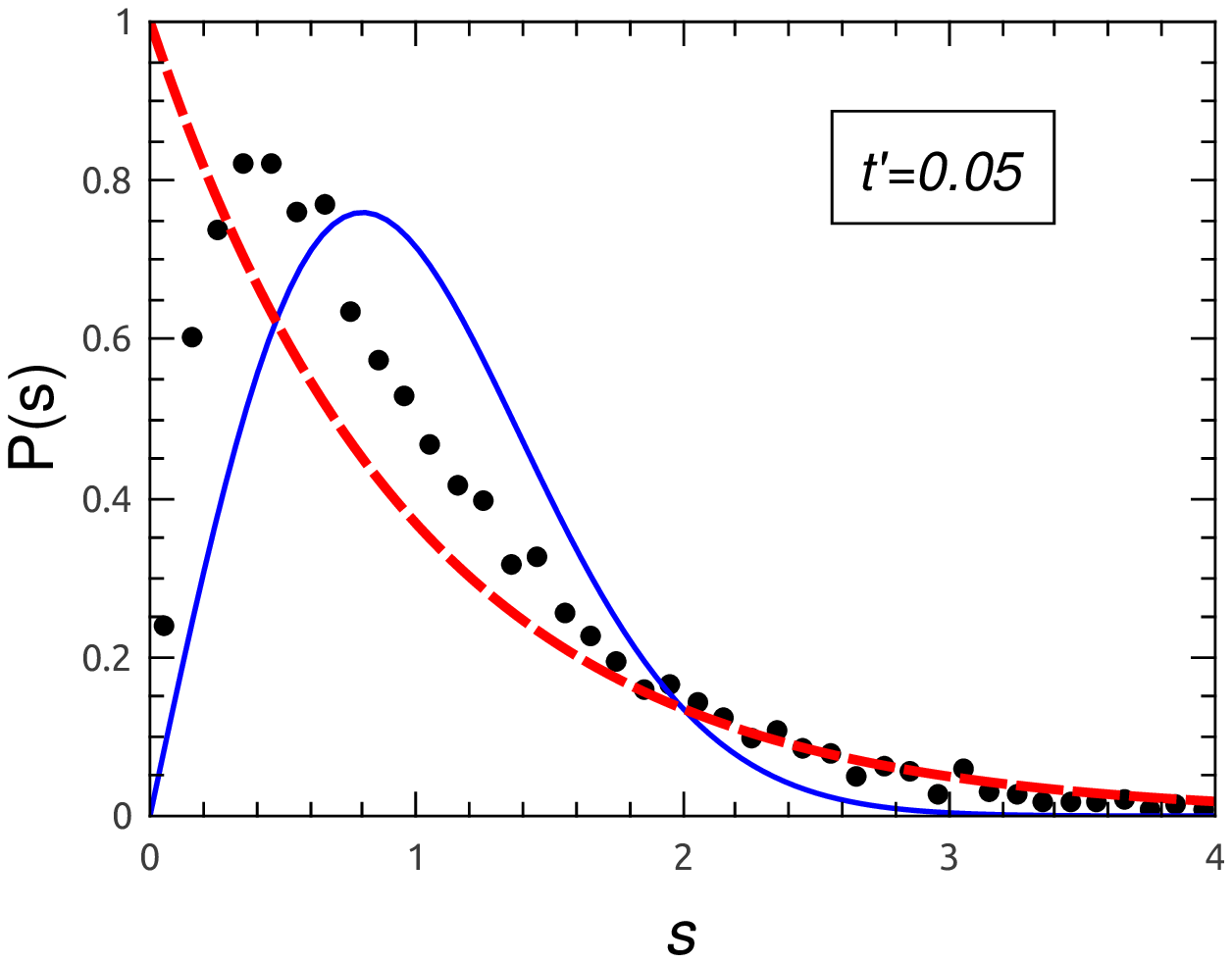} \\
\includegraphics[width=1.7in]{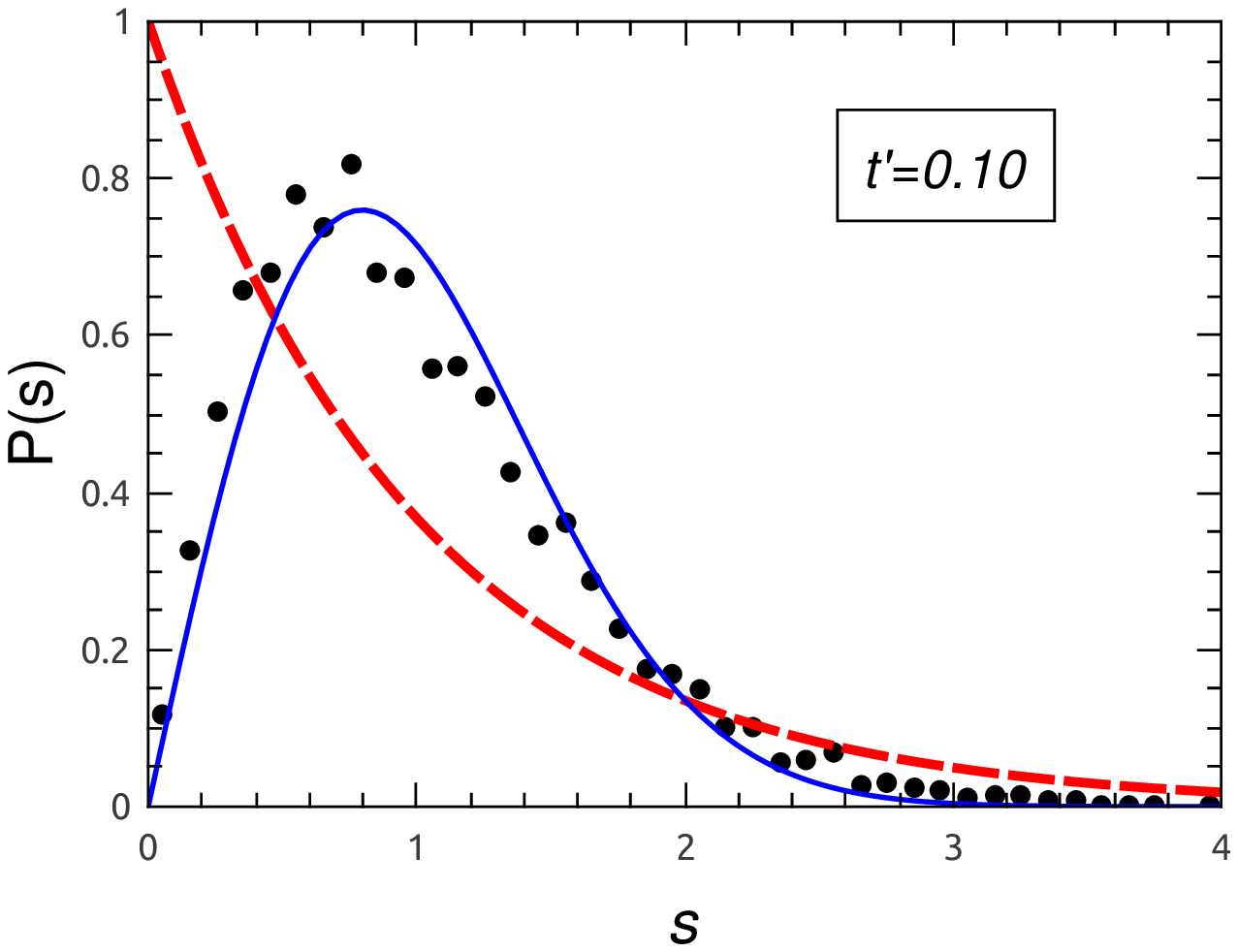}&
\includegraphics[width=1.7in]{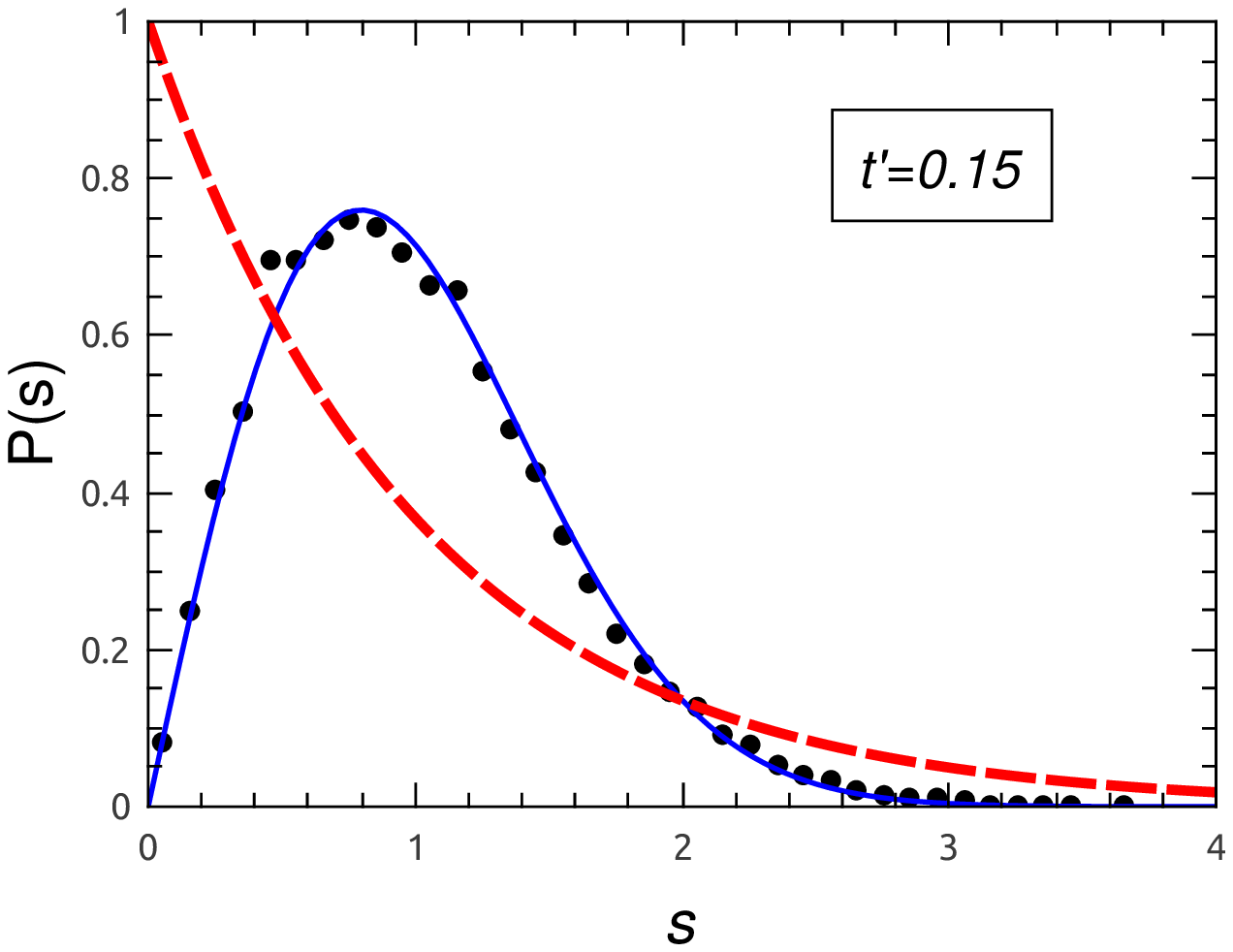}
\end{tabular}
\caption{(Color online) Level spacing distribution $P(s)$ for the Hubbard model
with $U=1$ at half filling  and $L=11$. The values of the integrability
breaking parameter $t'$ are 0.0,0.05,0.1,0.15.  As in
Fig.~\ref{Fig:levelspacing_tv}, the dashed line represents the Poisson
distribution and the solid line the GOE distribution.}
\label{Fig:levelspacing_hubbard}
\end{figure}

The level spacing distribution for our largest system sizes for the $t-t'-V$
model (Fig.~\ref{Fig:levelspacing_tv}) and the Hubbard
model(Fig.~\ref{Fig:levelspacing_hubbard}) for representative values of $V$ and
$U$ respectively show that $P(s)$ evolves from being Poissonian to GOE as the
integrability breaking parameter is increased. For the $t-t'-V$ model, we show
data for $V=1.0$ at half filling while increasing the integrability breaking
parameter $t'$ from 0. For the Hubbard model too we work at half filling,
setting $S_z=0(1)$ when the number of particles is even(odd). We show our
data for $U=1$ while increasing $t'$ from 0 in
Fig.~\ref{Fig:levelspacing_hubbard}.

\begin{figure}
\centering
\begin{tabular}{cc}
\includegraphics[width=1.7in]{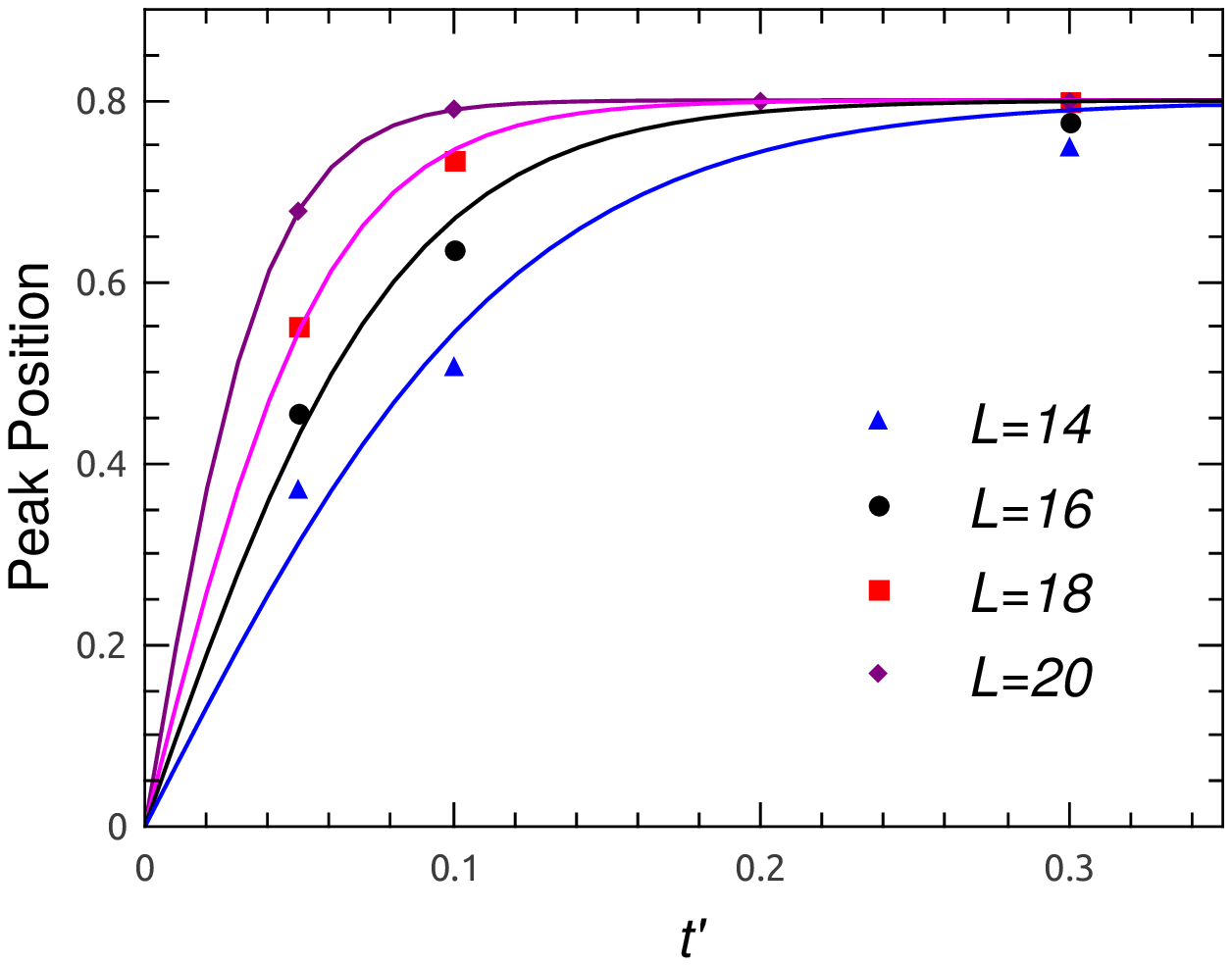} &
\includegraphics[width=1.7in]{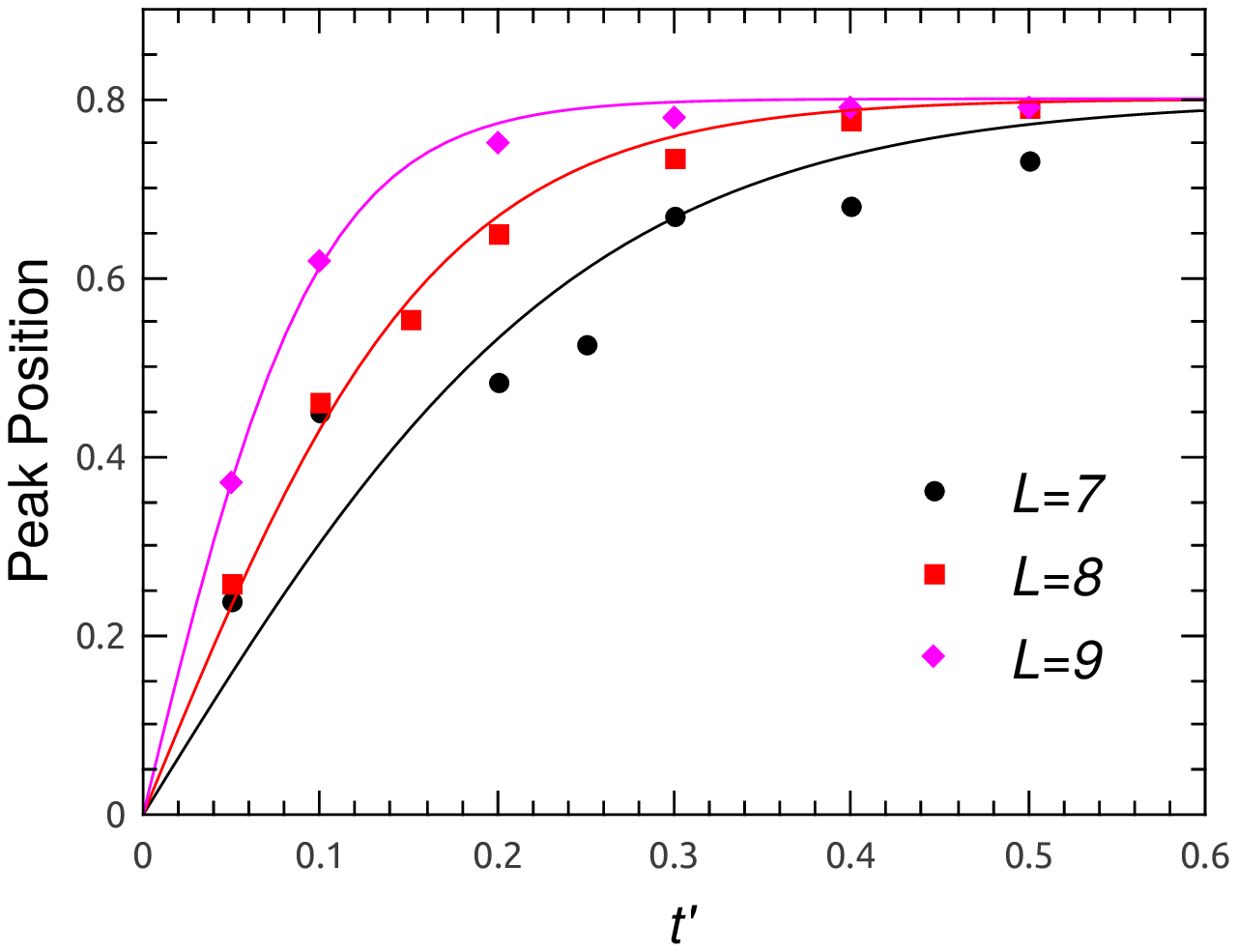}
\end{tabular}
\caption{(Color online) $(A)$ Peak position of $P(s)$ for the $t-t'-V$ model
with $V=2$ at half filling as a function on $t'$ for $L=14, 16, 18$ and $20$.
The solid lines are the function $0.8\tanh(t'/t'_{cr})$ used to obtain the
crossover scale of $t'$ as function of $L$ $(B)$ Peak position of $P(s)$ for
the Hubbard model  with $U=2$ as a function of $t'$ for $L=7, 8$ and $9$ to
obtain the crossover scale of $t'$ as a function of $L$. The solid lines are
the function $0.8\tanh(t'/t'_{cr})$. }
\label{Fig:peakposition_hubbard}
\end{figure}

Fig.~\ref{Fig:peakposition_hubbard} shows peak positions of $P(s)$ for
various $t'$ and $L$ obtained using the Brody distribution for
representative values $V=2.0$ and $U=2.0$ for the $t-t'-V$ and Hubbard models along
with fits to obtain the crossover scale for different $L$.

{\em Drude weight:} We assume that our system is in contact with an external
heat bath which causes it to thermalize at a temperature $T$ even if it is
integrable when isolated. One can then formally define a frequency
dependent charge
conductivity
\begin{equation}
\sigma_c (T, \omega) = D_c(T) \delta(\omega) + \sigma (\omega \neq 0, \omega),
\label{Eq:optcond}
\end{equation}
where $\omega$ is the frequency and $D_c(T)$ is the Drude weight or charge
stiffness~\cite{subroto.2008,rigol_shastry.2008,shastry.2006}. For an
integrable system without a charge gap, $D_c(T)$ can be argued to be non-zero
for all finite values of $T$ whereas for a non-integrable system it goes to
zero in the thermodynamic limit at any finite
temperature~\cite{subroto.2006,subroto.2008,rigol_shastry.2008}~\footnote{The
exception to this rule is a system which is supercondcuting. However, our
systems are one dimensional systems with short range interactions and cannot
display superconductivity at any finite $T$.}. Thus, $D_c(T)$ can be used as a
diagnostic tool to determine the crossover from integrability to
non-integrability. We consider the limit $T \rightarrow \infty$ for better statistics for
which~\cite{subroto.2006}
\begin{equation}
TD_c(T)=\frac{1}{LN}\sum_{\epsilon_n=\epsilon_m}|<n|J|m>|^{2},
\label{Eqn:charge_drudeweight}
\end{equation}
where $N$ is the size of the Hilbert space, $n$ and $m$ are energy eigenstates
at the same energy and $J$ is the charge current given by
\begin{eqnarray}
J=\lim_{k\to0}\frac{1}{k}[n(k),H].
\label{Eqn:charge_current}
\end{eqnarray}
Here $n(k)$ is the Fourier transform of the charge density. We emphasize that $D_c(T)$ is a useful
diagnostic tool only if the integrable system is gapless. For our specific
systems, this is true for the $t-t'-V$ model for all values of filling except
half filling for $V > 2|t|$, which we study later using only level spacing
statistics. For the Hubbard model, we calculate $D_c(T)$ away from half-filling, where it is
gapless.  At half filling, the integrable system has a charge gap but
no spin gap for all values of $U > 0$. In this case, the spin Drude weight
$D_s(T)$ can be used instead of $D_c(T)$. $D_s(T)$ can be obtained from
relations similar to~\ref{Eqn:charge_drudeweight} and~\ref{Eqn:charge_current}
with the charge current and density replaced by the spin current and
density respectively. We have verified that $D_s(T)$ yields the same
scaling of the integrability breaking parameter as $D_c(T)$.

\begin{figure}
\centering
\begin{tabular}{cc}
\includegraphics[width=1.7in]{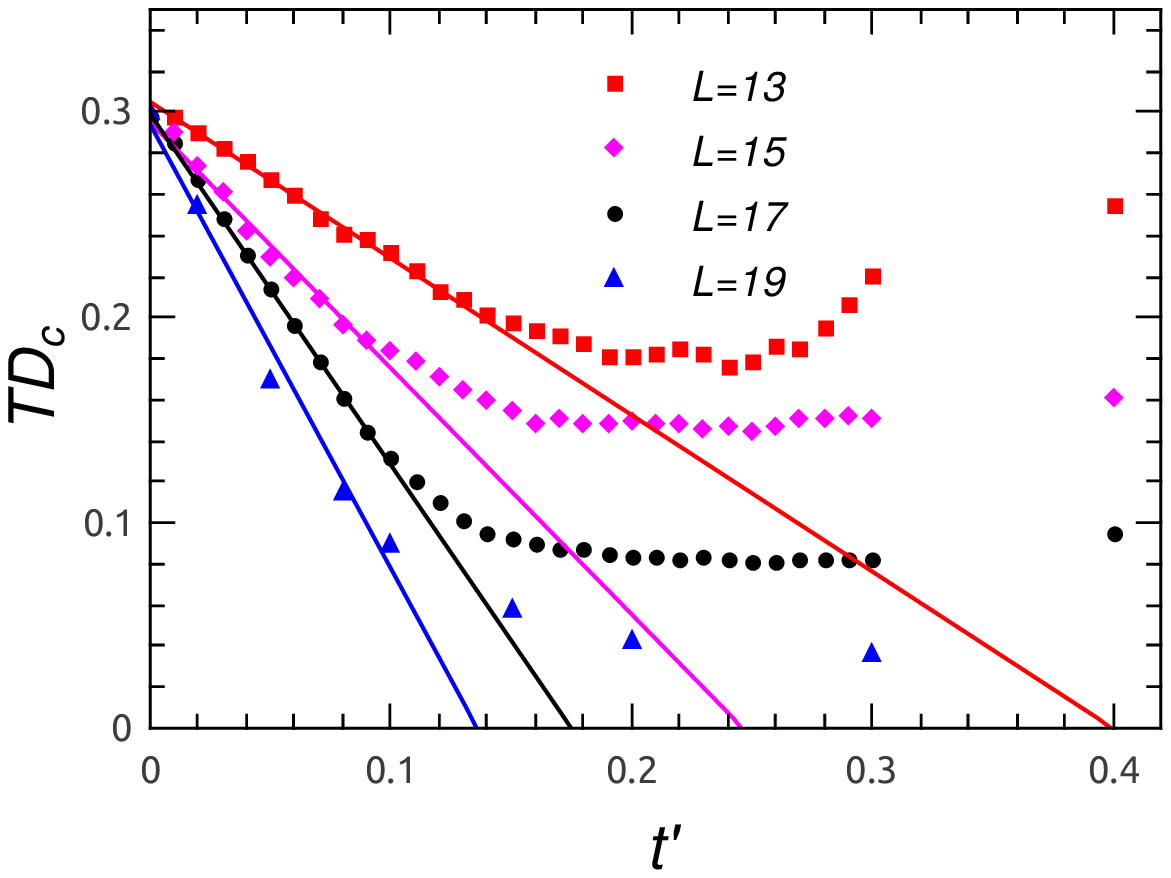}&
\includegraphics[width=1.7in]{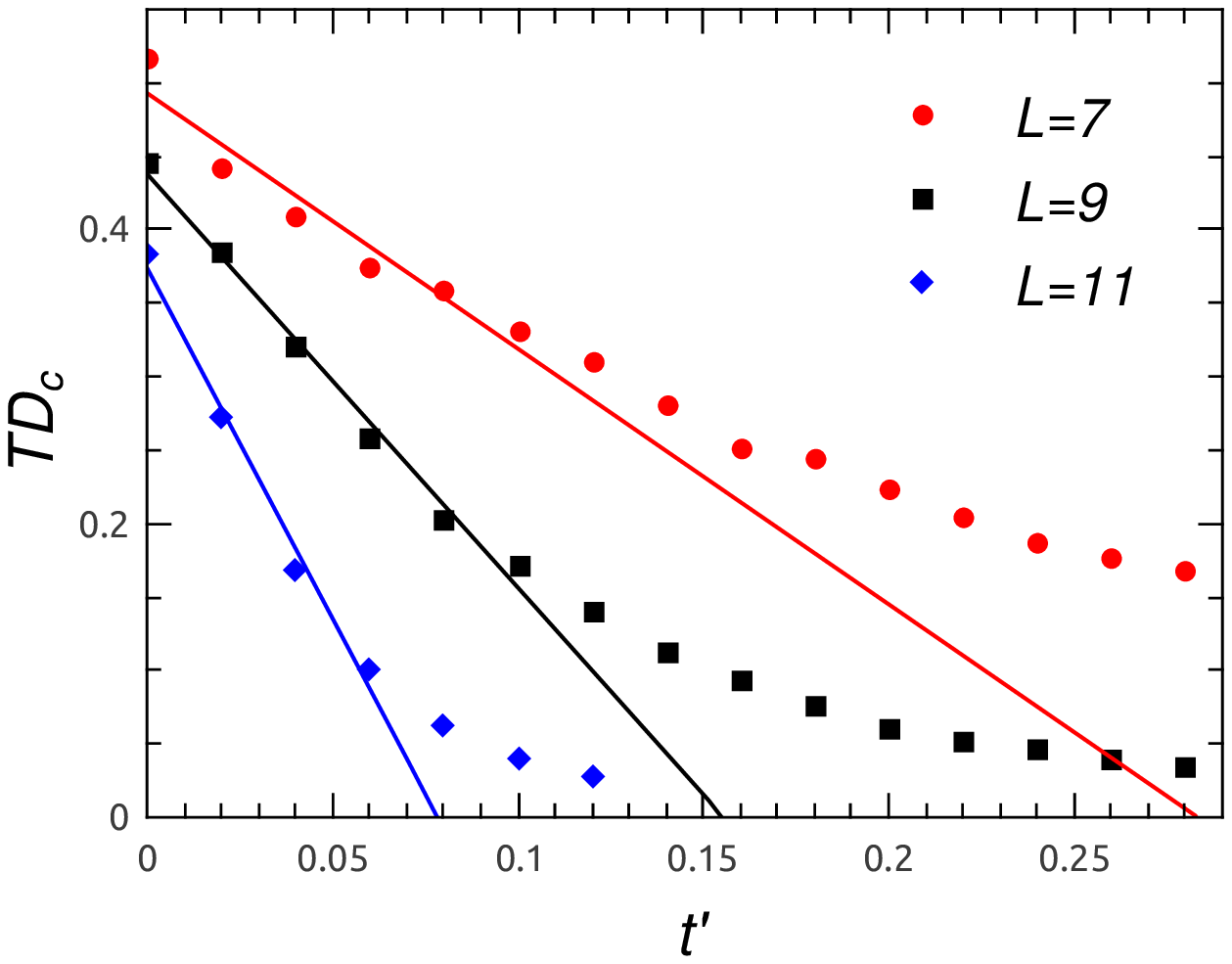}
\end{tabular}
\caption{(Color online) $(A)$ $TD_c(T)$ as $T\rightarrow \infty$ as a function
of $t'$ for the $t-t'-V$ model with $V=1$ for $L=13,15,17$ and $19$. The linear
fit to the data for small values of $t'$ is shown $(B)$ for the Hubbard model
with $U=2$ for $L=7,9$ and $11$. The linear fit for small values of $t'$ is
shown. }
\label{Fig:drude_tv_hubbard}
\end{figure}

Fig.~\ref{Fig:drude_tv_hubbard} shows $D_c(T)$ for the $t-t'-V$ model and the
Hubbard model. In the integrable limit of the $t-t'-V$ model ($t'=0$), $D_c(T)
= 0$ at half filling for an odd number of
particles~\cite{subroto.2008,herbrych.2011}.  In order to obtain a sufficient
number of data points for $D_c(T)$ at different values of $L$, we work away
from half-filling. For the Hubbard model we set $S_z=0$ always working with an
even number of particles even when away from half-filling.

We extract the value of the crossover scale of $t'$ for both models in the
following way: It can be seen from Fig.~\ref{Fig:drude_tv_hubbard} that
$TD_c(T)$ appears to decrease linearly for small values of $t'$ before leveling
off. Further, the value it appears to saturate to decreases with increasing
system size. We expect that in the thermodynamic limit, this value will be
equal to zero and will be attained for any non-zero value of $t'$. The
intercept on the $x$ axis of the linear fit at small values of $t'$ can thus be
used to define the crossover scale above which $TD_c(T)$ levels off. It is this
intercept that we determine as a function of system size $L$.

\begin{figure}
\centering
\begin{tabular}{cc}
\includegraphics[width=1.7in]{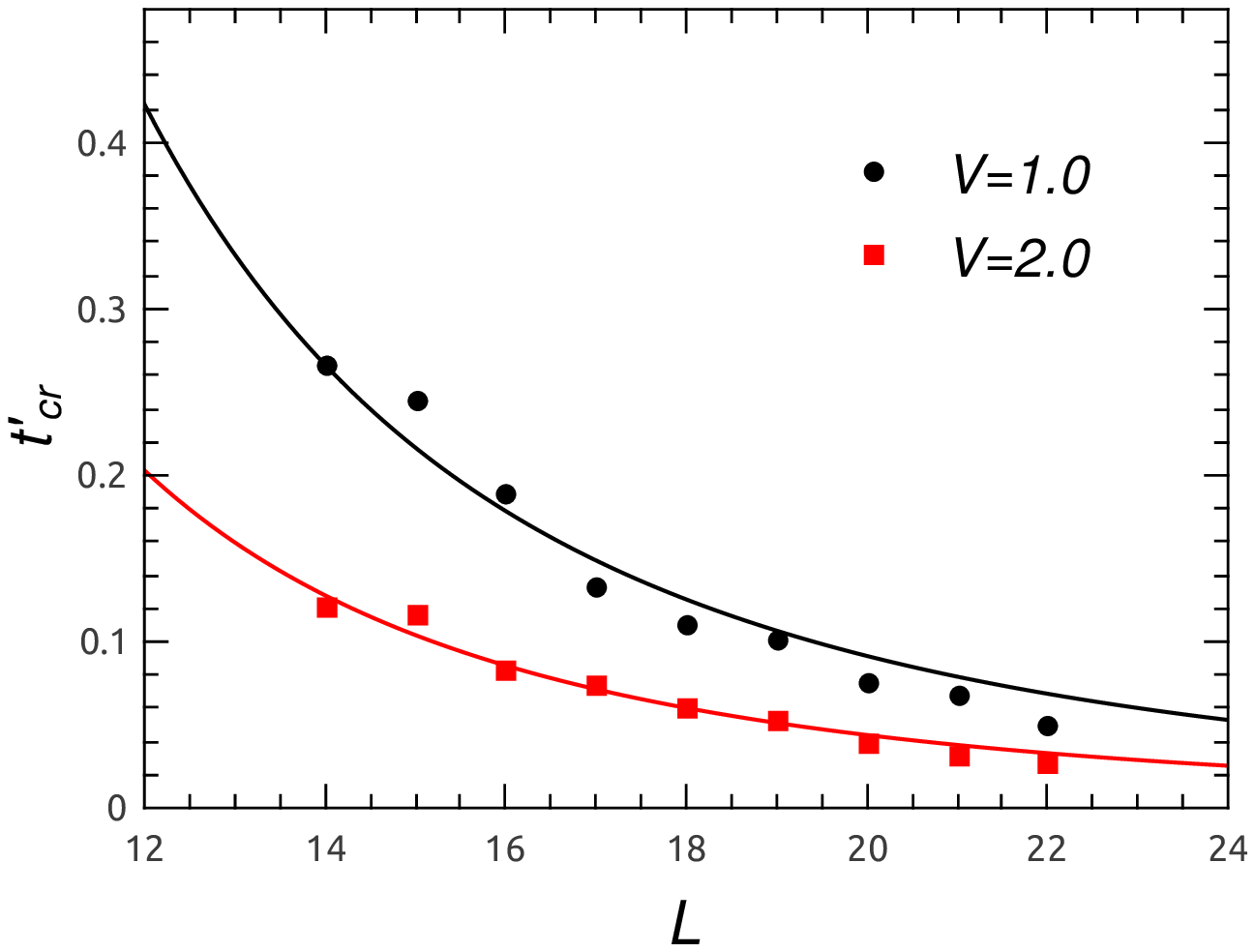}&
\includegraphics[width=1.7in]{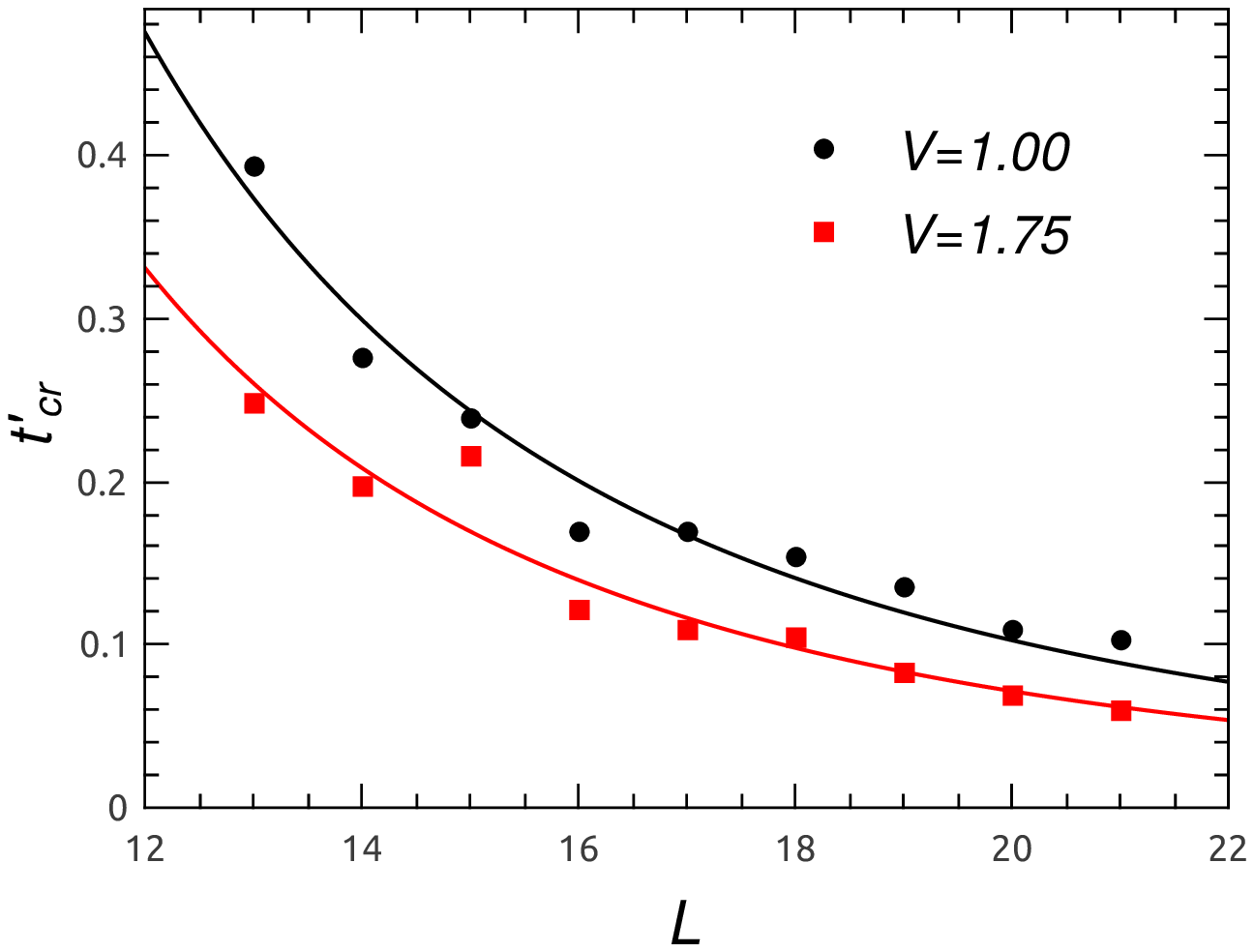}
\end{tabular}
\caption{(Color online)$t'_{cr}$ as function of $L$ for the $t-t'-V$ model with
$(A)$ $V=1$ and $V=2$ as obtained from the level spacing distribution and $(B)$
$t-t'-V$ model ($V=1$ and $V=1.75$) using the charge Drude weight. The solid
lines are fits to a power law decay given by $L^{-3}$}
\label{Fig:powerlaw tv}
\end{figure}

{\em Scaling of crossover scale with system size:} Fig.~\ref{Fig:powerlaw tv}
shows the scaling of the crossover value of the integrability breaking
parameter with system size as obtained from the level spacing distribution and
the charge Drude weight described above. It can be seen that values for a given system size are not identical owing to the fact
that they are obtained from two different methods and refer to a crossover
scale rather than a sharp threshold. However, what it is remarkable is that
they appear to scale in the same way with system size and the best fit to our
data shows that this scaling is $L^{-3}$. We show this scaling for two
different values of $V$ and have verified it for others as well.

\begin{figure}
\centering
\begin{tabular}{cc}
\includegraphics[width=1.7in]{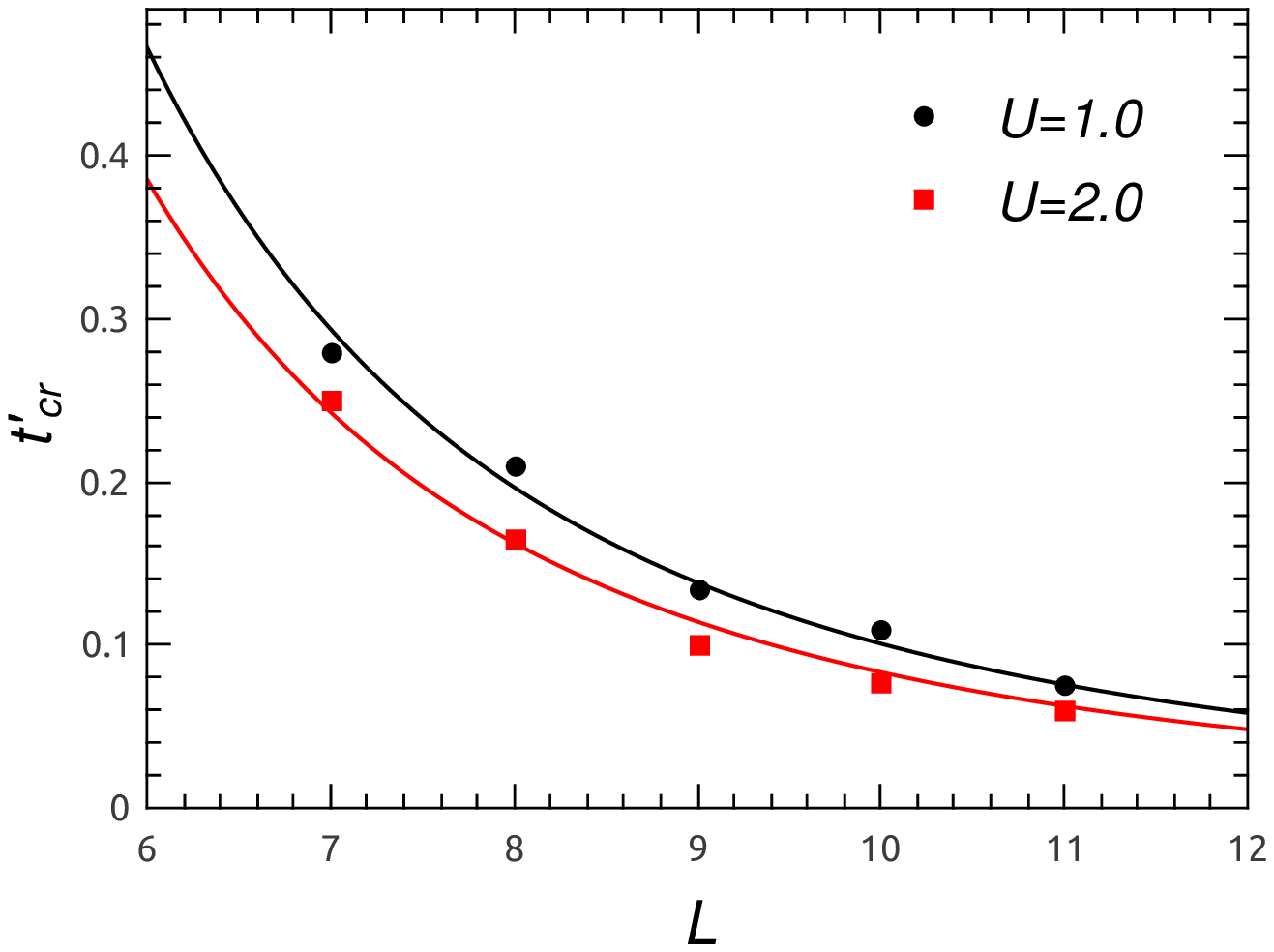}&
\includegraphics[width=1.7in]{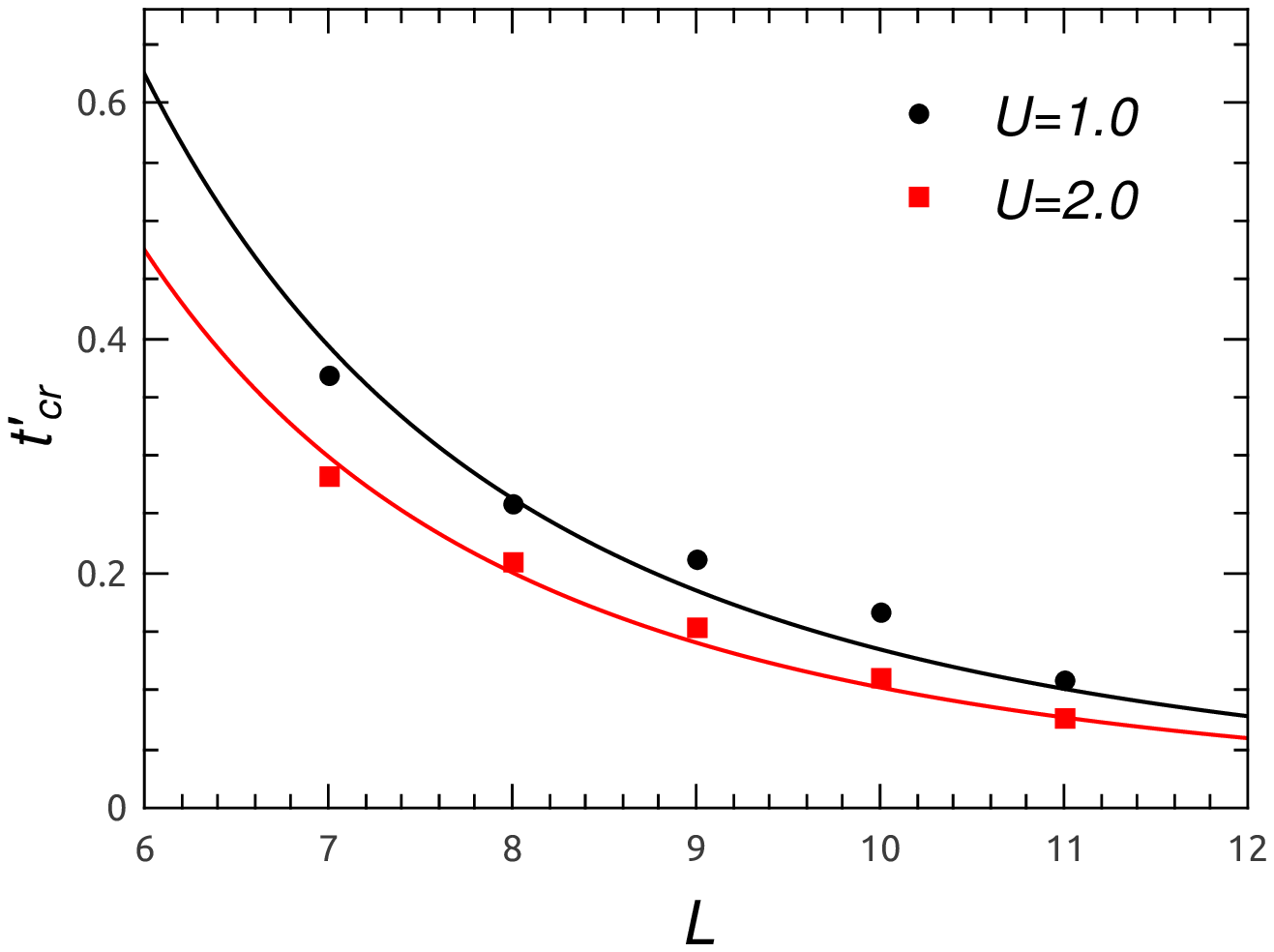}
\end{tabular}
\caption{(Color online)$t'_{cr}$ as function of $L$ for the Hubbard model with
$(A)$ ($U=1$ and $U=2$) as obtained from the level
spacing distribution and $(B)$ ($U=1$ and $U=2$) as obtained from the Drude weight  }
\label{Fig:powerlaw hub}
\end{figure}

Fig.~\ref{Fig:powerlaw hub} shows the crossover scale as a function of system
size obtained from the level spacing distribution and the Drude weight for the
Hubbard model for two different values of $U$. Once again, we see that the best fit
is of the form $L^{-3}$.

We have also verified numerically that the power law $L^{-3}$ is robust with
respect to the parameters in the Hamiltonian and the form of the
further neighbor integrability-breaking term.

Rabson {\em at. al.}~\cite{rabson.2004} had investigated a similar issue in one
dimensional spin models and had come to no definite conclusion about the
scaling of the crossover value of the integrability breaking parameter. Our
systems sizes are slightly larger than theirs, which enables us to make
better fits and we have verified to the extent possible that their
numerical data is consistent with the power law that we obtain.

\begin{figure}
\includegraphics[height=2.0in]{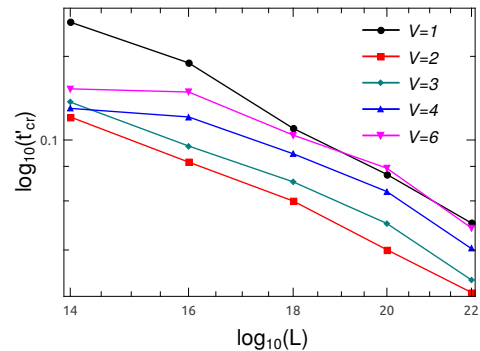}
\caption{(Color online)$t'_{cr}$ as function of $L$ for the $t-t'-V$ model for
$V=1,2,3,4$ and 6 as obtained from the level spacing distribution. The system
is gapless for $V=1$ and $2$ and gapped for $V=3,4$ and $V=6$ and it can be
seen that the $t'_{cr}$ for the three larger values of $V$ seems to be falling
off faster than a power law for the system sizes we have studied.}
\label{Fig:faster_power_law}
\end{figure}

We have also conducted a similar study for gapped systems using energy level statistics.
The specific system we have studied is the $t-t'-V$ model with $V > 2|t|$.
The data for integrability breaking crossover scale as a function of system
size is shown in Fig.~\ref{Fig:faster_power_law}. The value of $t'_{cr}$ as
obtained from energy level statistics is shown for $V=1$ and 2 for which the
integrable model is gapless and $V=3,4$ and 6 for which it is gapped. We have a
added a flux threading the loop since in its absence, one does not obtain the
right gap when there are an odd number of particles~\cite{rigol_gap.2010}. For the
system sizes we have studied, it appears that $t'_{cr}$ does falls off
faster than a power law when the system is gapped.

What is the origin of this power law? We do not have a very definite answer to
that question yet. However, on the evidence of our numerical data and the robustness
of the power law we conjecture that the exponent 3 is associated with the only
universal feature of the different models we study, namely the GOE ensemble
describing the non-integrable systems. If this is true, one will presumably
obtain different exponents when the non-integrable systems are described by
other ensembles. Our preliminary results on such microscopic models seems to bear out this fact and
a detailed study will be published later.

The fall off of the crossover scale faster than a power law when the system is
gapless is also intriguing. Naively, one might have expected the energy level
statistics, which is a property of the entire spectrum to not be affected by
the presence or absence of a gap. Our studies might suggest that in the systems
we study the entire specturm is controlled by the properties of a few low lying energy states.
A thorough validation of this claim
would require larger scale numerics, perhaps of the sort using the Density
Matrix Renormalization Group (DMRG) developed recently~\cite{karrasch.2012} or
tractable analytical calculations.

SR thanks R Nityananda for a discussion, and acknowledges support from a J C Bose Fellowship.
SM thanks Deepak Dhar for discussions and the Department of Science
and Technology, Government of India for support.
%
\end{document}